\documentclass[aps,showpacs,preprintnumbers,amsmath,amssymb,eqsecnum]{revtex4}

\usepackage{graphicx}
\usepackage{dcolumn}
\usepackage{bm}

\begin{document}

\newcommand{\balpha}{\bm{\alpha}}
\newcommand{\bmu}{\bm{\mu}}
\newcommand{\bsig}{\bm{\sigma}}
\newcommand{\brho}{\bm{\rho}}
\newcommand{\blam}{\bm{\lambda}}
\newcommand{\bL}{\bm{L}}
\newcommand{\bA}{\bm{A}}
\newcommand{\bB}{\bm{B}}
\newcommand{\bC}{\bm{C}}
\newcommand{\bS}{\bm{S}}
\newcommand{\br}{\bm{r}}
\newcommand{\bx}{\bm{x}}
\newcommand{\by}{\bm{y}}
\newcommand{\bz}{\bm{z}}
\newcommand{\bk}{\bm{k}}
\newcommand{\bv}{\bm{v}}
\newcommand{\mb}{\textbf m}
\newcommand{\pb}{\textbf p}
\newcommand{\rb}{\textbf r}
\newcommand{\sbd}{\textbf s}

\title{Estimates of isospin breaking contributions to baryon masses}

\author{Phuoc Ha}%
\email{pdha@towson.edu}
\affiliation{%
Department of Physics, Astronomy and Geosciences, Towson University, Towson MD 21252, USA
}%

\date{\today}

\begin{abstract}

We estimate the isospin breaking contributions to the baryon
masses which we analyzed recently using a loop expansion in the
heavy baryon approximation to chiral effective field theory. To
one loop, the isospin breaking corrections come from the effects
of the $d,\,u$ quark mass difference, the Coulomb and magnetic
moment interactions, and effective point interactions attributable
to color-magnetic effects. The addition of the first meson loop
corrections introduces new structure. We estimate the resulting
low-energy, long-range contributions to the mass splittings by
regularizing the loop integrals using connections to dynamical
models for finite-size baryons. We find that the resulting
contributions to the isospin breaking corrections are of the right
general size, have the correct sign pattern, and agree with the
experimental values within the margin of error.

\end{abstract}


\maketitle


\section{\label{sec:introduction} Introduction}

In a recent paper \cite{DH05}, we analyzed the structure of the
electromagnetic contributions to the mass splittings within
isospin multiplets in the baryon octet and decuplet.   In
particular, we studied the leading isospin breaking (IB)
contributions to the mass differences that come from the Coulomb
and magnetic interactions, quark mass differences, and the
one-loop mesonic corrections to those interactions.

Our analysis was based on  the standard chiral Lagrangian in the
heavy baryon approximation (HBA) to chiral effective field theory
as developed by Georgi \cite{Georgi_HBPT} and Jenkins and Manohar
\cite{Jenkins1}.   However, we used a spin- and flavor-index or
``quark'' representation of the effective octet and decuplet
baryon fields and the electromagnetic and mesonic interactions
rather than the usual matrix expressions for the fields.   The
connection of this representation to the usual effective-field
methods was discussed in detail in \cite{DHJ1,DHJ2}.  It has been
applied in past analyses of the splittings between baryon isospin
multiplets \cite{DHJ1,DHJ2,DH_masses}, and of the baryon magnetic
moments \cite{DHJ2,DH_moments3}. The results  in each case can be
summarized in terms of a set of effective interactions that have
the appearance of interactions between quarks in the familiar
semirelativistic or nonrelativistic quark models for the baryons.
The results are in fact completely general in their representation
of the relativistic heavy-baryon effective fields and their
interactions as shown in \cite{DHJ1,DHJ2}.  In the present case,
our results connect directly to Morpurgo's general parametrization
of the electromagnetic effects \cite{Morpurgo_EM1} derived from
QCD.

Our derivations in \cite{DH05} included the one-loop mesonic
corrections to the basic one-loop electromagnetic interactions, so
involved two loops overall. We derived a complete expression for
the isospin breaking electromagnetic contributions to the baryon
mass operator to this order, and then used the result to obtain
expressions for the operators that lead to intramultiplet mass
splittings.  As indicated, our emphasis was on the loop expansion,
rather than on the splittings obtained in a perturbative expansion
in powers of the chiral symmetry breaking parameters as in
\cite{Meissner}.   The Feynman loop integrals that appear in the
final results are of course the same as those obtained using the
usual effective fields, but have been reduced in \cite{DH05} for
loops containing heavy baryons to integrals for time-ordered
graphs.

 It remains to obtain explicit expressions for the mass splittings,
 calculate the integrals that appear, and compare the results with
 experiment, the subjects of the present paper.  We will do this by
 using the natural connection of the ``quark'' description of the
 effective fields  to semirelativistic dynamical models with the
 effective quarks playing the role of structure quarks.  This will
 allow us to model the structure of the matrix elements beyond the
 point-baryon approximation and estimate the matrix elements of
 the effective Hamiltonian for finite-size baryons.  The results
 will necessarily be model-dependent, but can be regarded as involving
 physically-motivated choices of cutoffs that emphasize the calculable
 long-distance parts of the loop corrections in the sense discussed by
 Donoghue, Holstein, and Borasoy \cite{Donoghue}.

The electromagnetic contributions to the baryon masses can be
expressed as a linear combination of the matrix elements of a set
of independent spin- and flavor-dependent operators $\Gamma_i$,
$(i=1,..., 32)$ in the baryon states \footnote{A complete set of
the operators $\Gamma_i$ is defined and analyzed in
\cite{DH05,Morpurgo_param}.}. There are  2 independent operators
that involve only one set of quark indices, called one-body
operators, 12 two-body operators, and 18 three-body operators.

We showed in \cite{DH05} that no three-body operators are
generated by one-loop mesonic corrections to the initial two-body
Coulomb and magnetic moment interactions after renormalization,
even though diagrams that connect three sets of quark indices are
initially present. The number of $\Gamma$'s that can appear to
this order is therefore reduced from 32 to 14. When the
calculation is restricted to the intramultiplet splittings, the
number of independent matrix elements is further reduced to 4
\cite{DH05,Morpurgo_param,Morpurgo_EM2}. As a result, we can bring
the electromagnetic mass-difference operator to the form
\begin{equation}
\label{Hib} \mathcal{H}_{\textrm{IB}} = a\sum_iM^d_i+ b\Gamma_4+
c\Gamma_5 +d\Gamma_{13} \, ,
\end{equation}
where
\begin{equation}
\Gamma_4=\frac{1}{2}\sum_{i\not = j}\,Q_iQ_j \, , \, \,
\Gamma_5=\frac{1}{2}\sum_{i\not = j}\,Q_iQ_j \bsig_i\cdot\bsig_j
\, , \, \, \Gamma_{13}=\frac{1}{2}\sum_{i\not = j}\,Q_iQ_jM_j^s \,
, \, \, \nonumber
\end{equation}
the $Q$s are quark charge matrices, and the matrices $M^d$ and
$M^s$ are defined as $M^d=\textrm{diag}\,(0,1,0)$,
$M^s=\textrm{diag}\,(0,0,1)$ \footnote{$M^s$ was denoted by $M$ in
\cite{DHJ1,DHJ2,DH_moments3} where the light-quark mass
differences did not play a role.}. The complete set of  one- and
two- body electromagnetic operators is given in Appendix
\ref{1-2-operators}.  The procedures needed to convert Eq.\
(\ref{Hib}) to an expression in terms of effective baryon fields
in the matrix representation are described in detail in
\cite{DHJ1,DHJ2}, but will not be needed here.

As will be discussed in Sec. \ref{sec:corrections}, the results in
\cite{DH05} omitted two further isospin breaking contributions
proportional to $m_d-m_s$ that are allowed by the general
effective field theory.  These are analogous to the effective
point interactions of the form $\sum\bsig_i\cdot\bsig_jM^s_i$ and
$\sum\bsig_i\cdot\bsig_jM^s_iM^s_j$ that appeared in our earlier
analysis of intermultiplet mass splittings  \cite{DHJ1,DHJ2}, but
involve the replacement of (one) matrix $M^s$ by $M^d$ and have
coefficients proportional to the mass difference $(m_d-m_u)$
rather than $m_s$. As shown below, terms of this form arise in
QCD-based dynamical models from a color-magnetic interaction
\cite{Sak-Zel}, and can give IB contributions to the
intramultiplet mass splittings comparable to those from
electromagnetic interactions \cite{Sakharov,FL1} \footnote{We
would like to thank Professor Jerrold Franklin for pointing out
the omission of the color-magnetic interaction in our paper
\cite{DH05}. We will use the label ``color-magnetic'' for these
terms, but emphasize that their presence is allowed in general.
Their coefficients can be treated as parameters as with the
analogous terms in \cite{DHJ1,DHJ2}.}.   Since the extra effective
interactions are two-body, the usual quark-model sum rules
\cite{Rubenstein1,Rubenstein2,Gal-Scheck} for baryon masses
continue to hold for matrix elements. Once again, only four
parameters $a$, $b$, $c$, and $d$ are needed to describe the
intramultiplet mass splittings through first order in $(m_d-m_u)$,
corresponding to an effective interaction of the form in Eq.\
(\ref{Hib}).

Our objectives in the present paper are, first, to extend the
theoretical analysis of isospin breaking mass splittings by
including the effects of the ``color-magnetic'' interactions
\cite{Sak-Zel} missed in \cite{DH05},  and, second, to estimate
 the final coefficients $a$, $b$, $c$, and $d$ in
$\mathcal{H}_{\textrm{IB}}$ numerically in the heavy baryon
approximation.   The first part of the analysis is completely
general.  The second will depend on specific models for baryon
structure.

As shown explicitly in Sec. \ref{coefficients}, the
electromagnetic contributions to the coefficients $a$, $b$, $c$,
and $d$ are given by combinations of cutoff-dependent Feynman
integrals over the three momenta in pion and photon loops.  The
effect of the soft, long-range parts of the interactions are
expected to be calculable.  We will therefore make reasonable
estimates of the soft parts of the integrals by employing known
form factors and dynamical models in cutting off the divergent
integrals. The differences from the experimental values give
estimates of the sizes of the residual hard, short-range
contributions.

With the color-magnetic interaction included, we find that the
calculated IB corrections are of the right general size and  have
the correct sign pattern to account  for the pattern of the
coefficients  in the IB mass-difference Hamiltonian. The
theoretical results agree with their experimental values at or
within the margin of error, especially given the uncertainty in
the theoretical input. Unless the corrections in the chiral
expansion from two meson loops (three loops overall) are larger
than we would expect, any deviations from the experiment can be
attributed to the presence of short-distance effects which can be
parametrized but not calculated in the effective field theory.



\section{\label{sec:corrections} Isospin breaking corrections to baryon masses}

It is necessary to recall that the leading IB corrections to
baryon masses come from the Coulomb interaction, magnetic moments
interactions, the effects of the $d,\,u$ quark mass difference,
and the color-magnetic interaction. We note that except for the
color-magnetic interaction, we have included the one-loop mesonic
corrections to other basic electromagnetic interactions (two loops
overall) in our calculations. We wish to evaluate the
contributions from these effects to the parameters $a$, $b$, $c$,
and $d$. For this goal, we firstly discuss the contributions from
the color-magnetic interaction to the mass splittings.

\subsection{\label{color-magnetic} Color-magnetic contributions}

The QCD color-magnetic interaction was first introduced by
Sakharov and Zel'dovich \cite{Sak-Zel} and further developed by De
R\'{u}jula, Georgi, and Glashow \cite{DGG} for their study on
hadron masses in a gauge theory. Sakharov \cite{Sakharov} and
Franklin and Lichtenberg \cite{FL1} showed that the QCD
color-magnetic interaction was of the same order of magnitude as
the purely electromagnetic interaction. The color-magnetic
contributions to the mass of a baryon $B$ arise in relativistic
quark models from the interaction
\begin{equation}
\label{Hcm}
\mathcal{H}_{\textrm{CM}}^B=\frac{4\pi\alpha_s}{9}\left[
\frac{\bsig_i\cdot\bsig_j}{m_im_j}\,\delta^3(\br_{ij})
+(j,k)+(k,i) \right] \, ,
\end{equation}
where $m_i$, $m_j$ represent effective quark masses, and
$\alpha_s$ is the strong coupling. At this point we will not
attempt to calculate the matrix elements of this interaction, but
will simply use its spin and mass structure to determine the
equivalent effective operators to use with the heavy-baryon
effective fields. We want, in particular,  to see the extent to
which the interaction in (Eq. (\ref{Hcm})) can be reduced to an
operator expression in terms of the $\Gamma$'s and $M^d$. For that
purpose, we take $m_u$ as the standard mass and consider the
symmetrical case in which $<\delta^3(\br_{ij})> \equiv A$, a
constant, for all states. We then substitute the identity
\footnote{$M^u=\textrm{diag}\,(1,0,0)$,
$M^d=\textrm{diag}\,(0,1,0)$, and $M^s=\textrm{diag}\,(0,0,1)$ are
the projection operators of the $u$-, $d$-, and $s$- quarks,
respectively.}
\begin{eqnarray}
\frac{1}{m_i}&=& \frac{1}{m_u}\left[M^u + \frac{m_u}{m_d}M^d + \frac{m_u}{m_s}M^s \right]_i  \nonumber \\
&=& \frac{1}{m_u}\left[\openone + \frac{m_u-m_d}{m_d}M^d +
\frac{m_u-m_s}{m_s}M^s\right]_i
\end{eqnarray}
into Eq. (\ref{Hcm}) and expand. To first order in $m_d-m_u$, the
color-magnetic interaction can then be rewritten as
\begin{equation}
\label{Hcmr} \mathcal{H}_{\textrm{CM}}^B=-C \sum_{i\not=j} \,
M_i^d \bsig_i\cdot\bsig_j + C'\sum_{i\not= j} M_i^dM_j^s
\bsig_i\cdot\bsig_j
 + \ldots \, ,
\end{equation}
where
\begin{equation}
C=\frac{4\pi\alpha_s}{9}\frac{A}{m_u^2}\frac{m_d-m_u}{m_u} \, , \,
\, \, C'=C\frac{m_s-m_u}{m_s} \, ,
\end{equation}
and the ellipsis represents terms that can be absorbed into the
structures that already appear in our analysis of intermultiplet
mass splittings in \cite{DHJ2}.

As mentioned above, the terms that are written out explicitly
involve isospin-breaking operators of order $m_d-m_u$ that are
allowed in the effective field theory \footnote{For example, in
heavy baryon effective field theory, the matrix form of the
operator $\sum M^d_i\bsig_i\cdot\bsig_j$ for the baryon octet is
$$
3{\rm Tr}\bar{B}\{M^d,B\} -{\rm Tr}\bar{B}[M^d,B] - 4{\rm Tr}(M^d)
{\rm Tr}(\bar{B}B) \, ,
$$
where $B$ is the usual matrix representation for the baryon
octet.} and could have been introduced from the beginning with
unknown coefficients to be determined from the data.  The
derivation suggests that these represent short-distance effects
not calculable in the chiral expansion.

We now show how to put the color-magnetic interaction in the
standard $a,b,c,d$ form. Using the identity
\begin{equation}
M^d=-3Q^2+\frac{4}{3}\openone-M^s \, ,
\end{equation}
we rewrite the operators in Eq. (\ref{Hcmr}) as
\begin{equation}
\sum_{i\not= j}M_i^d\bsig_i\cdot\bsig_j = \frac{4}{3}\sum_{i\not=
j}\bsig_i\cdot\bsig_j - \sum_{i\not=j}M_i^s \bsig_i\cdot\bsig_j
-6\Gamma_2,
\end{equation}
\begin{equation}
\sum_{i\not= j}M_i^dM_j^s\bsig_i\cdot\bsig_j =
\frac{4}{3}\sum_{i\not= j} M_j^s\bsig_i\cdot\bsig_j -
\sum_{i\not=j}M_i^s M_j^s\bsig_i\cdot\bsig_j -6\Gamma_{11}.
\end{equation}
The first two operators on the right in each expression contribute
to intermultiplet splittings but not to isospin splittings within
multiplets, so these can be dropped.
As far as the splittings are concerned, the following relations
hold
\begin{equation}
\Gamma_2=\frac{1}{3}\sum_iM_i^d+\frac{1}{2}(\Gamma_4-\Gamma_5) \,,
\, \, \, \Gamma_{11}=-\Gamma_{13}+\frac{1}{2}(\Gamma_4-\Gamma_5)
\, . \nonumber
\end{equation}
Hence, we find
\begin{equation}
\label{Hcmr1} \mathcal{H}_{\textrm{CM}}^B =
a_{\rm{CM}}\sum_iM^d_i+ b_{\rm{CM}}\Gamma_4+ c_{\rm{CM}}\Gamma_5
+d_{\rm{CM}}\Gamma_{13} \, ,
\end{equation}
where
\begin{equation}
\label{cm-contrib} a_{\rm{CM}}=-2C \, , \, \, b_{\rm{CM}}=3(C-C')
\, , \, \, c=-3(C-C') \, , \, \, d=6C' \, .
\end{equation}

Note that the sum rules for baryon masses continue to hold
\cite{Rubenstein1,Rubenstein2,Gal-Scheck}.

The one-loop mesonic corrections to the effective point
interactions do not introduce new isospin breaking terms, are
expected to be small, and will not be considered.


\subsection{\label{mass_splittings} Baryon mass splittings and sum rules}

The contributions of the operators $\sum_iM^d_i$, $\Gamma_4$,
$\Gamma_5$, and $\Gamma_{13}$ to the mass splittings within baryon
multiplets can be determined from the results given in
\cite{DH05,Morpurgo_param}. Using those results (see, for example,
Table I of the reference \cite{DH05} \footnote{Note that
$\Gamma_{10}$ and $\Gamma_{13}$ are too large by a factor of $2$
as given in Table I of \cite{DH05}. In addition, the old
evaluation of $\Gamma_{11}=-\Gamma_{14}$ was wrong.}) and Eq.\
(\ref{Hib}), we find
\begin{eqnarray} \label{splittings}
n - p & = & a - \frac{b}{3} - \frac{c}{3} \, , \nonumber \\
\Sigma^- - \Sigma^+ & = & 2a +  \frac{b}{3}- \frac{5c}{3} +
\frac{d}{3} \, , \nonumber \\
\Sigma^- - \Sigma^0 & = & a + \frac{2b}{3} - \frac{c}{3} + \frac{d}{6} \, , \nonumber \\
\Xi^- - \Xi^0 & = & a + \frac{2b}{3} - \frac{4c}{3} + \frac{d}{3}
\, , \nonumber \\
\Sigma^{*-} - \Sigma^{*+} & = & 2a + \frac{b}{3} + \frac{c}{3} +
\frac{d}{3}
\, , \\
\Sigma^{*-} - \Sigma^{*0} & = & a + \frac{2b}{3} + \frac{2c}{3} +
\frac{d}{6}
\, ,\nonumber  \\
\Xi^{*-} - \Xi^{*0} & = & a + \frac{2b}{3} + \frac{2c}{3} +
\frac{d}{3} \,
, \nonumber \\
\Delta^{++} - \Delta^0 & = & -2a + \frac{5b}{3} +  \frac{5c}{3} \, , \nonumber \\
\Delta^{++} - \Delta^- & = & -3a+b+c \, , \nonumber \\
\Delta^+ - \Delta^0 & = & -a + \frac{b}{3} + \frac{c}{3} \, .
\nonumber
\end{eqnarray}

Since there are only four independent parameters in
$\mathcal{H}_{\textrm{IB}}$, there are six sum rules among ten
mass differences. Below are the well-known sum rules
\cite{Coleman-Glashow,Rubenstein1,Rubenstein2,Gal-Scheck,Ishida}
\begin{eqnarray}
&& \Delta^0-\Delta^+ = n-p \, , \nonumber \\
&& \Delta^--\Delta^{++} = 3(n-p) \, , \nonumber \\
&& \Delta^0-\Delta^{++} = 2(n-p) + (\Sigma^0-\Sigma^+) - (\Sigma^--\Sigma^0) \, , \nonumber  \\
\label{sum_rules}
&& \Xi^--\Xi^0= (\Sigma^--\Sigma^+)-(n-p) \, , \\
&& \Xi^{*-}-\Xi^{*0} = (\Sigma^{*-}-\Sigma^{*+})-(n-p) \, , \nonumber \\
&& 2 \Sigma^{*0}-\Sigma^{*+} -\Sigma^{*-}= 2\Sigma^0-\Sigma^+-\Sigma^-. \nonumber
\end{eqnarray}
These sum rules hold for any set of purely one- and two-body
interactions as shown in
\cite{Rubenstein1,Rubenstein2,Gal-Scheck,Ishida}.

The sum rules can be violated by three-body operators. However, as
shown in \cite{DH05} and mentioned above, no effective three-body
operators are generated through one loop in the chiral expansion,
so any three-body effects must involve at least two meson loops
and are expected to be small. The sum rules are therefore expected
to hold with reasonable accuracy, as they do.

A weighted fit to the seven known mass splittings other than those
for the $\Delta$ baryons is given in Table
\ref{table:fit_to_data}.
\begin{table}[h]
\caption{\label{table:fit_to_data}A weighted fit to the seven
accurately known baryon mass splittings using the expressions in
Eq.\ (\protect{\ref{sum_rules}}). A best fit is obtained at values
(in MeV) of $a = 1.82 \pm 0.04$, $b = 3.35 \pm 0.24$, $c =-1.78
\pm 0.23$, and $d = 1.00 \pm 1.40$. The average deviation of the
fit from experiment is $0.13$ MeV. The experimental data are from
\cite{PDG}.}
\begin{ruledtabular}
\begin{tabular}{lrl}
Splittings & Calculated \, & Experiment \\
\hline
$n - p$ & 1.29 $\pm$ 0.12 & 1.293 $\pm$ 0.000 \\
$\Sigma^- - \Sigma^+$ & 8.05 $\pm$ 0.62 &  8.08 $\pm$ 0.08  \\
$\Sigma^- - \Sigma^0$ & 4.81 $\pm$ 0.30  &  4.807 $\pm$ 0.035\\
$\Xi^- - \Xi^0$ & 6.75 $\pm$ 0.58& 6.48 $\pm$ 0.24 \\
$\Sigma^{*-} - \Sigma^{*+}$ & 4.48 $\pm$ 0.49 & 4.40 $\pm$ 0.64 \\
$\Sigma^{*-} - \Sigma^{*0}$ & 3.03 $\pm$ 0.33 & 3.50 $\pm$ 1.12 \\
$\Xi^{*-} - \Xi^{*0}$ & 3.19 $\pm$ 0.52 & 3.20 $\pm$ 0.68  \\
$\Delta^{++} - \Delta^0$ & -1.02 $\pm$ 0.56 &  --- \\
$\Delta^{++} - \Delta^-$ & -3.88 $\pm$ 0.35 & --- \\
$\Delta^+ - \Delta^0$ & -1.29 $\pm$ 0.13 & --- \\
\end{tabular}
\end{ruledtabular}
\end{table}
A best fit is obtained at values (in MeV) of $a = 1.82 \pm 0.04$,
$b = 3.35 \pm 0.24$, $c =-1.78 \pm 0.23$, and $d = 1.00 \pm 1.40$
with an average deviation from experiment of 0.13 MeV and a
$\chi^2=1.67$ (with 7 degrees of freedom). Hereafter, we denote
the electromagnetic mass-difference operator with the best-fit
coefficients as $\mathcal{H}_{\textrm{IB}}^{\textrm{best}}$.

Using the data given in Table \ref{table:fit_to_data}, we find
that there are no significant violations of the sum rules.

\subsection{\label{coefficients} Expressions for the parameters $a$, $b$, $c$, and $d$ }

\begin{figure}
\caption{\label{Fig1} One-loop electromagnetic corrections to the
baryon mass due to the Coulomb interaction between quarks. Note
that the ``quarks" represent spin-flavor index sets on baryon
fields (not dynamical QCD quarks) and that spectator quarks are
suppressed.}
\includegraphics{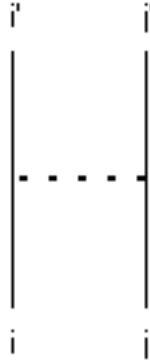}
\end{figure}

\begin{figure}[h]
\caption{\label{Fig2} Two-loop corrections to electromagnetic
interactions that involve meson exchange between quarks.
}
\includegraphics{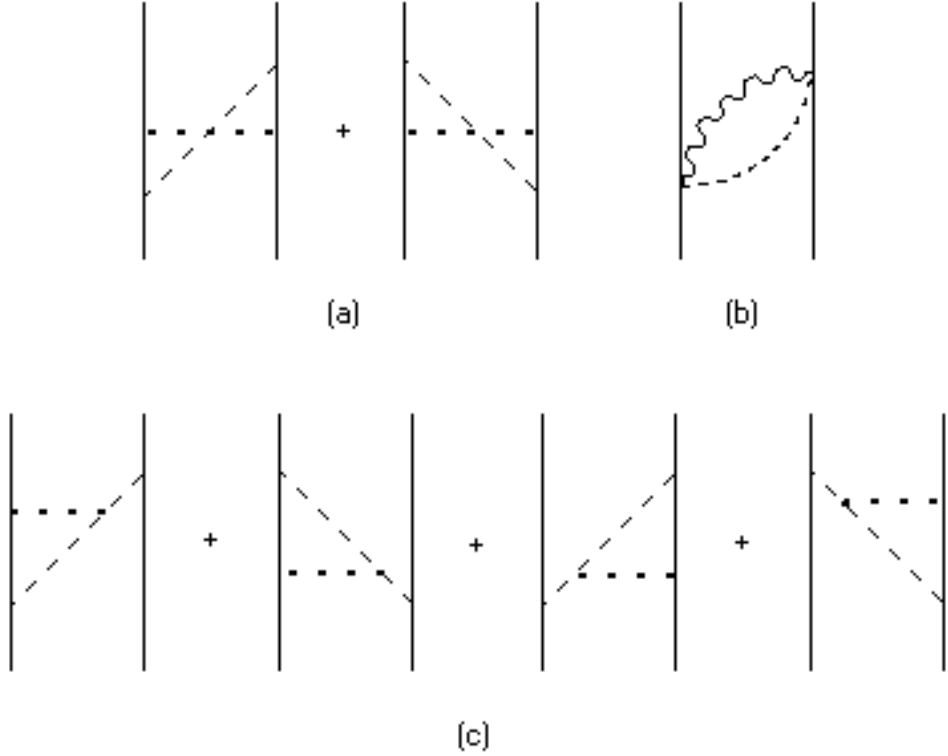}
\end{figure}

\begin{figure}[h]
\caption{\label{Fig3} (a): The basic meson exchange diagram. (b),
(c):  Electromagnetic contributions to the meson mass terms. (d):
electromagnetic correction to the  meson-quark vertex. }
\includegraphics{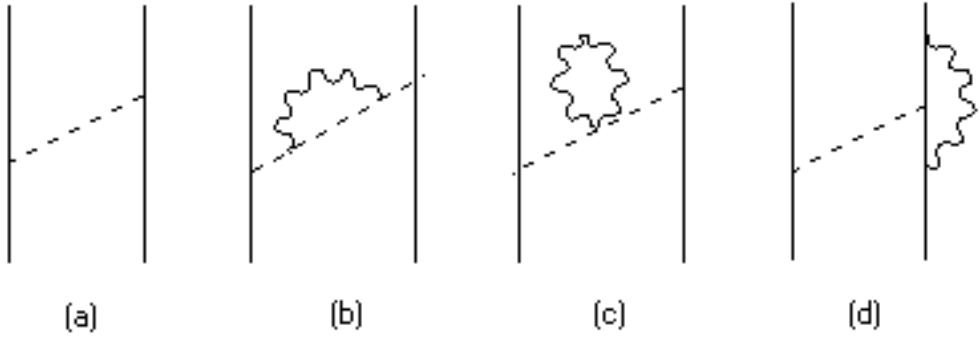}
\end{figure}

\begin{figure}[h]
\caption{\label{Fig4} Mesonic corrections to electromagnetic vertices.
}
\includegraphics{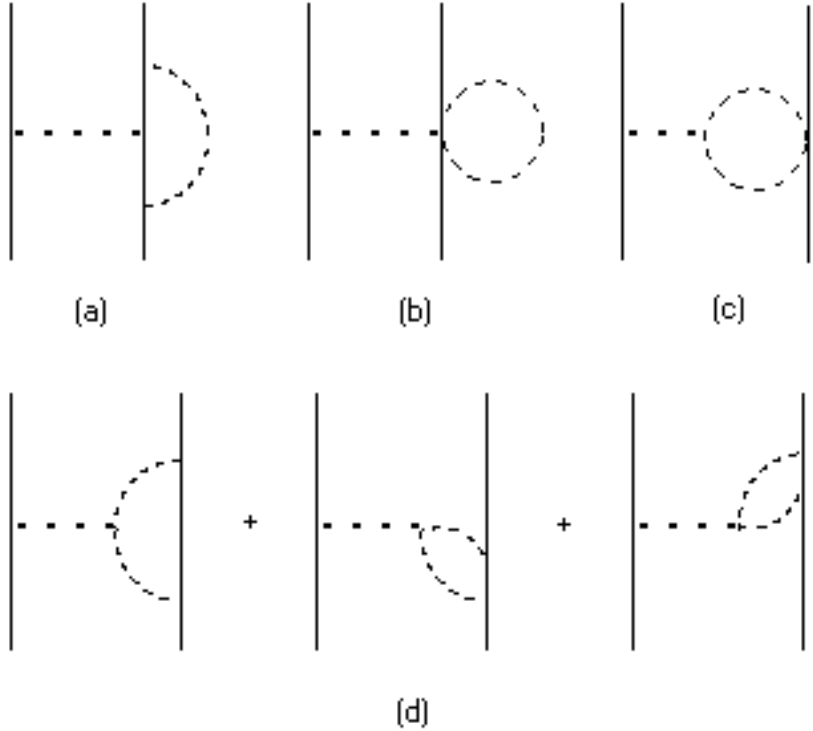}
\end{figure}

We are now ready to determine the expressions for the parameters
$a$, $b$, $c$, and $d$. Note that the IB mass-difference operator
$\mathcal{H}_{\textrm{IB}}$ can be written as \footnote{In
\cite{DH05}, $\mathcal{H}_1$, $\mathcal{H}_2$, and $\mathcal{H}_3$
are denoted by $\mathcal{H}_{\textrm{charge}}$,
$\mathcal{H}_{\textrm{moment}}$, and $\mathcal{H}_{du}$,
respectively.}
\begin{equation}
\label{H1a} \mathcal{H}_{\textrm{IB}} = \mathcal{H}_1 +
\mathcal{H}_2+ \mathcal{H}_3 + \mathcal{H}_{\textrm{CM}} \ .
\end{equation}
The first term in this expression, $\mathcal{H}_1$, is the total
contribution to the baryon mass differences from charge
interactions
\begin{eqnarray}
\mathcal{H}_1 &=& [I_{QQ}+6\mathcal{I}_{1,\pi}-
8(2\mathcal{I}_{2,\pi}+\mathcal{I}_{2,K})]\Gamma_4
 - 2\mathcal{I}_{1,\pi}(\Gamma_5-\Gamma_2)  \nonumber \\
&& + 24(\mathcal{I}_{2,\pi}-\mathcal{I}_{2,K})\Gamma_{13} -
2(\mathcal{I}_{1,\pi}-\mathcal{I}_{1,K})(6\Gamma_{13} -
3\Gamma_{10}-2\Gamma_{14}+\Gamma_{11})
\label{Hem} \\
&& +\left[2\mathcal{I}_{1,\pi}-\frac{8}{3}( \mathcal{I}_{2,\pi}
 -\mathcal{I}_{2,K})\right](M^d_i+M^d_j+M^d_k) \ , \nonumber
\end{eqnarray}
where \footnote{An extra term,
$-12(\mathcal{I}_{1,\pi}-\mathcal{I}_{1,K})$, in the coefficient
of $\Gamma_{13}$ at the second row of Eq. (4.16) in \cite{DH05} is
now deleted.  Note also that in \cite{DH05}, $\mathcal{I}_{2,l}$
is missing a factor $\frac{1}{2}$ and there is an extra factor
$4\pi$ in the definition of $I_{\mu\mu}$.}
\begin{equation}
\label{II1} \mathcal{I}_{1,l }=
-\frac{1}{3}I_{1,l}+\frac{2}{3}I_{2,l}+
\frac{2}{3}I_{3,l}-\frac{2}{3}I_{4,l} \ ,
\, \, \, \, \,
\mathcal{I}_{2,l} = \frac{1}{2}(I_{5,l}+I_{6,l}-I_{7,l}-I_{8,l}) \ ,
\end{equation}
In these expressions $I_{QQ}$ is an integral associated with the
Coulomb interaction diagram in Fig.\ \ref{Fig1}, and $I_{i,l}$
$(i=1,...,8; \, l=\pi, K, \eta)$ are the integrals associated with
the  diagrams shown in Figs. \ref{Fig2}(a) - \ref{Fig2}(c),
\ref{Fig3}, \ref{Fig4}(a) - \ref{Fig4}(d), respectively, that
contribute to the baryon mass differences through charge
interactions \footnote{Here $I_4$ and later $I_{\mu\mu}$, $I_9$,
and $I_{10}$ are redefined so that the integrals are all in MeV.}.

\begin{figure}[h]
\caption{\label{Fig5} Instantaneous magnetic moment-moment
interactions and mesonic corrections. A zigzag line with crosses
at the vertices represents a factor $\mathcal{H}_{\mu\mu} \equiv -
\mu_i\mu_j\bsig_i\cdot\bsig_j\,I_{\mu\mu}$  }
\includegraphics{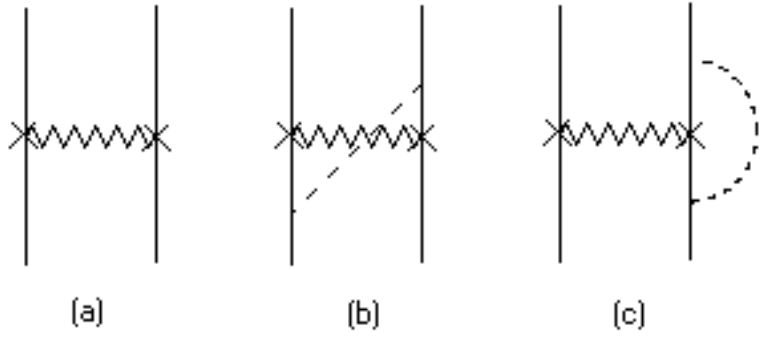}
\end{figure}

As discussed in \cite{DH05}, we work in the heavy baryon limit,
and the original Feynman  integrals reduce to integrals over three
momenta in old fashioned time-ordered perturbation theory. Thus,
in Figs.\ \ref{Fig1}-\ref{Fig5}, a solid vertical line represents
a quark moving upwards toward later times, dashed lines represent
mesons, wiggly lines represent transverse photons, and a
horizontal dotted line represents the instantaneous Coulomb
interaction between the particles on which it terminates. Only
quark lines involved in the interactions and representative time
orderings are shown. The incoming and outgoing quark lines are to
be collected into the corresponding baryons so that, viewed at the
baryon level of the usual baryon chiral perturbation theory,
Figs.\ \ref{Fig1} and \ref{Fig3}(a) are one-loop diagrams while
Figs.\ \ref{Fig2}, \ref{Fig3}(b-d), and \ref{Fig4} are two-loop
diagrams. Note that our Figs.\ \ref{Fig1}-\ref{Fig5} are identical
to Figs. 2(b), 4, 5, 7, and 8 in \cite{DH05}, respectively. The
evaluation of the various integrals is discussed in the next
Section.


The second term in Eq. (\ref{H1a}), $\mathcal{H}_2$, is the total contribution
to the baryon mass differences from magnetic moment interactions
\begin{eqnarray}
\mathcal{H}_2 &=& -I_{\mu\mu}(\mu_a^2\Gamma_5 + 2\mu_a\mu_b\Gamma_{14}) + \mathcal{H}'_{9}+ \mathcal{H}'_{10} \nonumber \\
\label{H_moment}
&& + \frac{2}{9}\left[\mu_a(9\mu_a+\mu_b)I_{9,\pi}
-2\mu_a(\mu_a+\mu_b)\left(I_{9,\pi}-I_{9,K}\right)\right]\sum_iM^d_i,
\end{eqnarray}
where
\begin{eqnarray}
\mathcal{H}'_{9} & = & -\frac{2}{3}\mu_a^2 I_{9,\pi} (\Gamma_2 + 3 \Gamma_4)+
   6 \mu_a^2 (I_{9,\pi} - \frac{4}{27}I_{9,\eta}) \Gamma_5 \nonumber \\
&& + \frac{2}{3}\mu_a^2 (I_{9,\pi} - I_{9,K}) (9 \Gamma_{10} + \Gamma_{11})
  + 4 \mu_a \left[ \mu_a I_{9,\pi} - (\mu_a+\mu_b)I_{9,K}\right] \Gamma_{13} \nonumber \\
&& + \frac{4}{3} \mu_a \left[ - 9 \mu_a I_{9,\pi} + 5(\mu_a+\mu_b)I_{9,K}
     + \frac{4}{3} (3\mu_a+2\mu_b) I_{9,\eta} \right] \Gamma_{14} \ ,
\end{eqnarray}
and
\begin{eqnarray}
\mathcal{H}'_{10} & = & \frac{4}{3} \mu_a^2 \,(4 I_{9,\pi} +
3 I_{9,K} + \frac{2}{3}I_{9,\eta}) \,\Gamma_5 \nonumber \\
& & + \frac{4}{3} \mu_a \left[ (5 \mu_b-3\mu_a) I_{9,\pi} + (\mu_a +7 \mu_b)I_{9,K}
+ 2(\mu_a + \frac{5}{3}\mu_b)I_{9,\eta} \right] \,\Gamma_{14} \ .
\end{eqnarray}
Here, $\mu_a=2.793$ , $\mu_b=-0.933$; $I_{\mu\mu}$ and $I_{9,l}$
are the integrals associated with the direct interaction between
magnetic moments (Fig. \ref{Fig5}(a)) and the moment-moment
interaction including one-loop mesonic corrections (Figs.
\ref{Fig5}(b) and \ref{Fig5}(c)), respectively.

The third term in Eq. (\ref{H1a}), $\mathcal{H}_3$, involving the
effects of the $d,\,u$ quark mass differences on the baryon masses
and on the single meson exchange amplitude, is of the form
\begin{equation}
\label{Hdu}
\mathcal{H}_3 = \Delta_{du}\sum_iM^d_i + \frac{2\Delta^M_q}{3 \Delta^M_{\textrm{em}}}I_{4,K^0} \,
\left[ 12\Gamma_{13}-6\Gamma_{10}-4\Gamma_{14}+2\Gamma_{11}\right] \, ,
\end{equation}
where $\Delta^ M_q  \equiv M_{K^0}^2 - M_{K^\pm}^2 +
\Delta^M_{\textrm{em}}$, $\Delta^M_{\textrm{em}}=
M_{\pi^\pm}^2-M_{\pi^0}^2$, and
$\Delta_{du}=[(m_d-m_u)/(m_s-m_u)]\tilde{\alpha}_m$ is originally
the coefficient of $\sum_iM^d_i$ from the single-particle mass
operator at the quark level \footnote{ See Eq. (2.38) in
\cite{DH05}.}. Using the value $\tilde{\alpha}_m\approx 178$ MeV
obtained in the absence of loop corrections \cite{DHJ2,DH_masses}
and  the ratio  $(m_d-m_u)/(m_s-m_u)\approx 0.227$ \cite{Weinberg}
gives the estimate $\Delta_{du} \approx 4.04$ MeV. Here, we will
take into account the possible loop corrections to
$\tilde{\alpha}_m$ and will treat $\Delta_{du}$ as a parameter.

The operators $\Gamma_i$ that appear above are all independent.
However, contributions of their matrix elements to intramultiplet
mass splittings satisfy a number of relations with
$\Gamma_{10}=-\Gamma_{13}$,
$\Gamma_{11}=-\Gamma_{14}=\Gamma_{10}+\frac{1}{2}(\Gamma_4-\Gamma_5)$,
and
$\Gamma_2=-\frac{1}{3}\sum_iM^d_i+\frac{1}{2}(\Gamma_4-\Gamma_5)$.
Note also that $\Gamma_{19}=\Gamma_{20}$ and
$\Gamma_{25}=\Gamma_{26}$ do not contribute. We can therefore
bring $\mathcal{H}_j$, $(j=1,2,3)$ to a form similar to Eq.
(\ref{Hib})
\begin{equation}
\label{H_j}
\mathcal{H}_j = a_j\sum_iM^d_i+b_j\Gamma_4+c_j\Gamma_5+d_j\Gamma_{13} \, ,
\end{equation}
where the coefficients of $\mathcal{H}_1$ are
\begin{eqnarray}
\label{coeff_1}
a_1 &=& \frac{4}{3}[\mathcal{I}_{1,\pi}-2(\mathcal{I}_{2,\pi}-\mathcal{I}_{2,K})] \, , \nonumber \\
b_1 &=& I_{QQ}+4\mathcal{I}_{1,\pi}+3\mathcal{I}_{1,K}-8(2\mathcal{I}_{2,\pi}+\mathcal{I}_{2,K}) \, ,  \nonumber \\
c_1 &=& -3\mathcal{I}_{1,K} \, , \\
d_1 &=& 12 [-(\mathcal{I}_{1,\pi}-\mathcal{I}_{1,K})
 + 2(\mathcal{I}_{2,\pi}-\mathcal{I}_{2,K})] \, , \nonumber
\end{eqnarray}
the coefficients of $\mathcal{H}_2$ are
\begin{eqnarray}
\label{coeff_2}
a_2 &=& \frac{2}{9} \mu_a \left[ (8 \mu_a-\mu_b) I_{9,\pi} + 2(\mu_a + \mu_b)I_{9,K} \right] \, , \nonumber \\
b_2 &=& \mu_a\mu_bI_{\mu\mu}+
\frac{1}{3}\mu_a\left[(18\mu_a-10\mu_b)I_{9,\pi}-
(13\mu_a+24\mu_b)I_{9,K}-12(\mu_a + \mu_b)I_{9,\eta}\right] \, ,  \nonumber \\
c_2 &=& -\mu_a(\mu_a+\mu_b)I_{\mu\mu}+
\frac{1}{3}\mu_a\left[10(\mu_a+\mu_b)I_{9,\pi}+
(25\mu_a+24\mu_b)I_{9,K}+12(\mu_a + \mu_b)I_{9,\eta}\right] \, , \\
d_2 &=& -2\mu_a\mu_b I_{\mu\mu} + \frac{4}{3}\mu_a \left[ (5\mu_b
- 14\mu_a)I_{9,\pi} +(8\mu_a + 9\mu_b)I_{9,K} + 6(\mu_a +
\mu_b)I_{9,\eta} \right] \, , \nonumber
\end{eqnarray}
and, finally, the coefficients of $\mathcal{H}_3$ are
\begin{equation}
\label{coeff_3} a_3 = \Delta_{du}\, , \ \ b_3 = 2 (\Delta^ M_q
/\Delta^M_{\textrm{em}}) I_{4,K^0} \, , \ \ c_3 = -2 (\Delta^ M_q
/\Delta^M_{\textrm{em}}) I_{4,K^0} \, , \ \ d_3 = 8 (\Delta^ M_q
/\Delta^M_{\textrm{em}}) I_{4,K^0} \, .
\end{equation}
It follows from Eqs. (\ref{Hib}), (\ref{H1a}), and (\ref{H_j})
that
\begin{equation}
\label{coeff}
a = \sum_{i=1}^3 a_i + a_{\textrm{CM}} \, , \ \ b = \sum_{i=1}^3 b_i + b_{\textrm{CM}}\, ,
\ \ c = \sum_{i=1}^3 c_i + c_{\textrm{CM}}\, , \ \ d = \sum_{i=1}^3 d_i + d_{\textrm{CM}}\, .
\end{equation}
The coefficients $a_{\textrm{CM}}$, $b_{\textrm{CM}}$,
$c_{\textrm{CM}}$, and $d_{\textrm{CM}}$ are defined by Eq.
(\ref{cm-contrib}).

In the next section, we will first evaluate the integrals and then
study how well the dynamical theory developed in our earlier
analyses of the baryon masses and magnetic moments describes the
coefficients in $\mathcal{H}_{\textrm{IB}}$.


\section{\label{sec:integrals} The integrals}

Let us start with the Coulomb integral $I_{QQ}$ and the magnetic
integral $I_{\mu\mu}$ defined as
\begin{equation}
\label{Coulomb_int1}
I_{QQ} = e^2 \int\frac{d^3k}{(2\pi)^3|\bk|^2} \, , \,\,\,
I_{\mu\mu} = \frac{2\mu_N^2}{3}\int\frac{d^3k}{(2\pi)^3} \, ,
\end{equation}
where $\mu_N=e/(2m_N)$ is the nucleon magneton.

Note that Eq. (\ref{Coulomb_int1}) refers only to the ``bare"
Coulomb and magnetic interactions between quarks without effects
of baryon structure. These integrals and others to follow are
formally divergent.  The divergences are commonly absorbed in the
HBA to chiral effective field theory into unknown constants
representing short-distance effects that are to be evaluated from
experiment.

In dynamical models, the integrals arise from matrix elements of
the corresponding operators in physical baryon states, and those
matrix elements involve additional momentum-dependent factors
associated with the structure of the baryon which may regularize
the integrals \footnote{Recall that the ``quark'' lines in, for
example, Fig.\ 1, keep track of the spin and flavor content of the
baryons in our representation, but do not introduce separate
momentum factors in Feynman integrals. Only the energy and
momentum of the entire baryon and of exchanged photons or mesons
are relevant as discussed earlier \cite{DHJ1,DHJ2,DH05}.}. Our
objective here is to estimate the soft or long-distance
contributions to the integrals, and to see the extent to which
these account for the magnitude and pattern of baryon mass
splittings. We therefore adopt the dynamical approach. This
requires information on the internal structure of the baryons,
some of which is available through measured baryon form factors
and the semirelativistic theory of baryon structure, but its use
necessarily involves models that go beyond the HBA to chiral
effective field theory.

The semirelativistic theory of
baryon structure has been considered by a number of authors and is
quite successful \cite{Brambilla,Carlson,Capstick}. For
simplicity, we will use the model considered in \cite{DH_moments3}
in which the baryon masses are calculated variationally for the
semirelativistic Hamiltonian of Brambilla \textit{et al.} using
Gaussian wave functions. The results agree with those of a similar
calculation by Carlson, Kogut, and Pandharipande \cite{Carlson}
and are consistent with those of the much more extensive
calculations of Capstick and Isgur \cite{Capstick}.

We will use Jacobi coordinates to describe the positions of the
quarks. Define
\begin{eqnarray}
{\rb}_{ij} &=& \bx_i - \bx_j \, , \ \
{\bf R}_{ij} = \frac{m_i\bx_i + m_j\bx_j}{m_{ij}} \, , \nonumber \\
\label{Jacobi_space}
{\rb}_{ij,k} &=& {\bf R}_{ij} - \bx_k \, = \,
\frac {m_i(\bx_i - \bx_k) + m_j(\bx_j - \bx_k)}{m_{ij}} \, , \nonumber \\
{\bf R}_{ijk} &=& \frac {m_{ij}{\bf R}_{ij} + m_k\bx_k}{M} \, ,
\end{eqnarray}
where the $\bx_i$ are the particle coordinates, $m_{ij}=m_i+m_j$,
$M=m_i+m_j+m_k$, and ${\bf R}_{ijk}$ is the usual center-of-mass
coordinate. The roles of $i$, $j$, and $k$ are completely
symmetric at this stage. However, it is reasonable to neglect the
very small difference between the effective masses of the $u$ and
$d$ quarks in the dynamical calculations. At least two of the
quarks in each baryon are then identical or have the same mass. We
label these 1 and 2, with the odd quark labelled 3 and then define
the internal Jacobi coordinates  $\brho$ and $\blam$ as $\brho =
{\rb}_{12}$ and $\blam = {\rb}_{12,3}$. Alternatively, we can use
coordinates with the role of $(1,2)$ replaced by $(2,3)$ or
$(3,1)$ in the definition, and define $\brho' = {\rb}_{23}$,
$\blam' = {\rb}_{23,1}$, or $\brho'' = {\rb}_{31}$, $\blam'' =
{\rb}_{31,2}$. The coordinate pairs $\brho'$, $\blam'$ and
$\brho''$, $\blam''$ can be expressed in terms of $\brho$ and
$\blam$ and conversely, so one can work with whichever of the
pairs is most convenient and switch between them as necessary. The
spatial volume element is simply $d^3R\,d^3\rho\,d^3\lambda$, and
is equivalent for the other pairs of internal coordinates.

In the position space, we can express $I_{QQ}$ and $I_{\mu\mu}$,
for the symmetrical case \footnote{In general, the Coulomb
integral $I_{QQ}$ is a function of both variational parameters
$\beta_\rho$ and $\beta_\lambda$ that have different values for
different baryons. However, as discussed in \cite{DH05}, the
baryon dependence of $\beta_\rho$ and $\beta_\lambda$ can be
ignored.}, as follows
\begin{equation}
\label{Coulomb_int2}
I_{QQ}= \frac{e^2}{4\pi}\int d^3\rho\, d^3\lambda\, \frac{\left|\psi(\brho,\blam)\right|^2}{\rho} , \, \, \,
I_{\mu\mu}= \frac{2\mu_N^2}{3}\int d^3\rho\, d^3\lambda\, \left|\psi(\brho,\blam)\right|^2 \delta^3(\brho) \ .
\end{equation}
Using the position-space variational Gaussian wave functions in
\cite{DH_moments3} for the $L=0$ ground states
\begin{equation}
\label{gausswf}
\psi_0(\brho,\blam) = \left( \frac {\beta_\rho \beta_\lambda }{ \pi} \right)^{3/2}
 \exp{ [ -\frac{1}{2}(\beta_\rho^2 \rho^2 + \beta_\lambda^2 \lambda^2)]} \ ,
\end{equation}
and changing coordinates appropriately \cite{DH_moments3},
we can easily calculate $I_{QQ}$ and $I_{\mu\mu}$. The results are
\begin{equation}
\label{IqqImm}
     I_{QQ} = \frac{2\alpha_{\textrm{em}}}{\sqrt{\pi}} \beta_\rho , \, \, \,
     I_{\mu\mu} = \frac{2\mu_N^2}{3} \frac{\beta_\rho^3}{\sqrt{\pi^3}} ,
\end{equation}
where $\alpha_{\textrm{em}}=e^2/4\pi$. For $\beta_\rho=340$ MeV
obtained for the nucleon, $I_{QQ}=2.80 \, \textrm{MeV}$ and
$I_{\mu\mu}=0.123 \, \textrm{MeV}$.

Next, we consider the integrals associated with the mesonic
corrections.


$I_{1,l}$ comes from the diagram in Fig. \ref{Fig2}(a) and, as
shown in \cite{DH05}, factors into the product of a Coulomb
integral and a mesonic integral $I_{1,l}'$.
\begin{equation}
\label{I1l_2} I_{1,l} = I_{QQ} \times I_{1,l}' \, , \, \, \,
I_{1,l}'=
\frac{\beta^2}{4f^2}\int\frac{d^3k'}{(2\pi)^32E_l'}\frac{k^{'2}}{E_l^{'2}}F_A^2(\bk^{'2})
\, ,
\end{equation}
where $\beta$ is a common dynamical matrix element for meson
emission \cite{DHJ2} \footnote{In \cite{DH05}, $1/f$ is to be
replaced with $\beta/f$.}, $E_l(\bk')=\sqrt{\bk^{'2}+M_l^2}$, and
$F_A(\bk^{'2})$ is the axial vector form factor of the baryon
introduced when we neglect the excited states and include the
internal structure of the baryon through the baryon wave function.

To evaluate the mesonic integral $I_{1,l}'$, we use the dipole
form factor $F_A(\bk^{'2})=\Lambda_A^4/(\Lambda_A^2+\bk^{'2})^2$
used in our earlier analyses of baryon masses
\cite{DH_masses,DHJ2}. The mesonic integral $I_{1,l}'$ is
convergent and easy to be numerically calculated. For $\Lambda_A
=850$ MeV and $\beta=0.75$, we find that $I_{1,\pi}'=0.039$,
$I_{1,K}'=0.016$, and $I_{1,\eta}'=0.014$.


$I_{2,l}$ comes from the diagram in Fig. \ref{Fig2}(b) and is
given by \footnote{The coefficient of $I_{2,l}$ in \cite{DH05} is
off by a factor of $4$.}
\begin{equation}
\label{I2l} I_{2,l} =  \frac{e^2\beta^2}{f^2}
\int\frac{d^3k}{(2\pi)^3|\bk|} \int
\frac{d^3k'}{(2\pi)^32E'_l}\frac{1}{E+E'_l}(F_Q(\bk^2)F_A(\bk^{'2}))^2
\, .
\end{equation}
Note that a product of the form factors, $F_Q(\bk^2)F_A(\bk^{'2})$
is introduced at each vertex of the diagram to ensure that
$I_{2,l}$ is convergent. To evaluate the integral numerically, we
use a nucleon form factor
$F_Q(\bk^2)=\Lambda_Q^4/(\Lambda_Q^2+\bk^2)^2$ with $\Lambda_Q
=843$ MeV \cite{nucleon_FFs}.

$I_{3,l}$ comes from the diagram in Fig. \ref{Fig2}(c) whose
intermediate matrix element, showing all quarks, is like a
baryon-meson scattering matrix element. We will introduce a
product of charge form factors $F_Q(\bk^2)F_M(\bk^2)$ to the
intermediate state where
$F_M(\bk^2)=\Lambda_M^4/(\Lambda_M^2+\bk^2)^2$ is the meson charge
form factor with $\Lambda_M =1017$ MeV. Hence,
\begin{equation}
\label{I3l} I_{3,l} = \frac{e^2 \beta^2}{4f^2}
\int\frac{d^3k}{(2\pi)^3|\bk|^2}F_Q(\bk^2)F_M(\bk^2) \int
\frac{d^3k'}{(2\pi)^3}
\frac{\bk'\cdot\bk''}{4E_l'E_l''}\frac{E_l'+E_l''}{E_l'E_l''}F_A(\bk^{'2})F_A(\bk^{''2})
\, ,
\end{equation}
where $\bk''=\bk'+\bk$ .


$I_{4,l}$ comes from the diagram in Fig. \ref{Fig3}(a)
differentiated with respect to $M_l^2$. It is defined as
\begin{equation}
\label{I4l} I_{4,l} =   \frac{\beta^2
\Delta^M_{\textrm{em}}}{4f^2} \int \frac{d^3k'}{(2\pi)^32E_l'}
\frac{k^{'2}}{E_l'^3} F_A^2(\bk^{'2}) \, .
\end{equation}
This integral can be analytically evaluated. The result is
\begin{equation}
\label{I4l_result} I_{4,l} = \frac{\beta^2
\Delta^M_{\textrm{em}}}{512 \pi f^2}
\frac{\lambda_A^5(M_l+5\lambda_A)}{(M_l+\lambda_A)^5} \, .
\end{equation}
%

$I_{5,l}$ is the integral associated with the diagram in Fig. 4(a)
and is identical to $I_{1,l}$, i.e., $I_{5,l}= I_{1,l}$.


$I_{6,l}$ comes from the diagram in Fig. \ref{Fig4}(b). The
extended structure at the vertex in Fig. \ref{Fig4}(b) suggests
that the same form factor as in Fig. \ref{Fig4}(a) should be used.
For the rest of the diagram, a Coulomb interaction must be
absorbed by the wave function (not the form factor). So, for the
symmetrical case, we get
\begin{equation}
\label{I6l} I_{6,l} = \frac{e^2\beta^2}{4f^2}
\int\frac{d^3k}{(2\pi)^3|\bk|^2} \int d^3\lambda\, \left|\tilde
\psi(\bk,\blam)\right|^2 \int \frac{d^3k'}{(2\pi)^32E'_l}
F_A^2(\bk^{'2}) \, ,
\end{equation}
where
\begin{equation}
\label{tpsi}
\left|\tilde \psi(\bk,\blam)\right|^2=
\int d^3\rho\ e^{-i \bk \cdot \brho} \left|\psi(\brho,\blam)\right|^2 \, .
\end{equation}

Using the Gaussian wave functions, Eq. (\ref{gausswf}), we find
\begin{equation}
\int d^3\lambda\, \left|\tilde \psi(\bk,\blam)\right|^2 = \exp [- k^2/(4 \beta_\rho^2)]  \, , \, \, \,
I_{QQ}=e^2 \int\frac{d^3k}{(2\pi)^3|\bk|^2} e^{- k^2/(4 \beta_\rho^2)} \, .
\end{equation}
Hence,
\begin{equation}
\label{I6ls} I_{6,l} = I_{QQ} \times
\frac{\beta^2}{4f^2}\int\frac{d^3k'}{(2\pi)^32E_l'}F_A^2(\bk^{'2})
\ .
\end{equation}
%


$I_{7,l}$ and $I_{8,l}$ come from the diagrams in Figs.
\ref{Fig4}(c) and \ref{Fig4}(d), respectively. The intermediate
state of the diagram in Fig. \ref{Fig4}(d) has the baryon-meson
scattering structure, thus it involves $F_Q(\bk^2)F_M(\bk^2)$. The
diagram in Fig. \ref{Fig4}(c) is related to the one in Fig.
\ref{Fig4}(d). Its intermediate part involves the baryon-meson
scattering structure contracted with a meson-meson-baryon vertex
but instantaneous, not scattering. Putting in the form factors,
\begin{equation}
\label{I7l} I_{7,l} = \frac{e^2\beta^2}{4f^2}
\int\frac{d^3k}{(2\pi)^3|\bk|^2} \int d^3\lambda\, \left|\tilde
\psi(\bk,\blam)\right|^2 \int
\frac{d^3k'}{(2\pi)^3}\frac{1}{E'_l+E''_l}F_A(\bk^{'2})F_A(\bk^{''2})
\, ,
\end{equation}
and
\begin{equation}
\label{I8l} I_{8,l} = \frac{e^2 \beta^2}{4f^2}
\int\frac{d^3k}{(2\pi)^3|\bk|^2}F_Q(\bk^2)F_M(\bk^2) \int
\frac{d^3k'}{(2\pi)^3}\frac{\bk'\cdot\bk''}{E_l'E_l''}
\frac{1}{E_l'+E_l''}F_A(\bk^{'2})F_A(\bk^{''2}) \, .
\end{equation}

At this point, it is important to note that the diagrams in Fig.
\ref{Fig4} are related by electromagnetic current conservation.
The sum of these diagrams associated with mesonic corrections to
the photon-quark vertex must give a coefficient for the $1/\bk^2$
Coulomb singularity that vanishes in the limit of zero photon
momentum, $\bk\rightarrow 0$. This condition is found to hold for
the diagrams without form factors \cite{DH05}. It is
straightforward to check that the condition still hold for the
diagrams with wave functions and form factors. Indeed, since the
integrals $I_{6,l}$ and $I_{7,l}$ appear with opposite-sign
coefficients and for $\bk\rightarrow 0$ the integral over
$\lambda$ in their expressions is just the normalization integral
and hence approaches unity, the coefficient of $1/\bk^2$ vanishes
when $I_{6,l}$ and $I_{7,l}$ are combined. Similarly, we can
easily show the cancellation between the $I_{5,l}$ and $I_{8,l}$
terms for $\bk\rightarrow 0$ if we write $I_{5,l}$ explicitly and
notice that the form factors in the first factor in $I_{8,l}$ also
approach unity for $\bk\rightarrow 0$.

Finally, $I_{9,l}$ and $I_{10,l}$ come from Figs. \ref{Fig5}(b)
and \ref{Fig5}(c), respectively. They are identical and factor
into the product of $I_{\mu\mu}$ and the mesonic integral
$I_{1,l}'$
\begin{equation}
\label{I9l}
I_{9,l} = I_{10,l}=I_{\mu\mu} \times I_{1,l}' \, .
\end{equation}

It is straightforward to evaluate $I_{1,l}$,$I_{2,l}$,$I_{6,l}$,
and $I_{9,l}$ numerically. To evaluate $I_{3,l}$, $I_{7,l}$, and
$I_{8,l}$, we integrate first on $d\Omega_{k'}$ with the polar
axis chosen along $\bk$. The obtained results are then integrated
numerically.

We present in Table\ \ref{table:integrals} the numerical values of
$I_i$ $(i=1, ..., 9)$, $\mathcal{I}_1$, and $\mathcal{I}_2$ for
different meson loops. All the integrals are measured in MeV.
Calculations use $\alpha_{\textrm{em}}=1/137$, $f \approx 93.0$
MeV, $\beta=0.75$, $\lambda_A=850$ MeV, $\lambda_Q=843$ MeV,
$\lambda_M=1017$ MeV, and $\beta_\rho=340$ MeV.

\begin{table}
\caption{Numerical values of $I_i$ $(i=1, ..., 9)$,
$\mathcal{I}_1$, and $\mathcal{I}_2$ for different meson loops.
All the integrals are measured in MeV. Calculations use
$\alpha_{\textrm{em}}=1/137$, $f \approx 93.0$ MeV, $\beta=0.75$,
$\lambda_A=850$ MeV, $\lambda_Q=843$ MeV, $\lambda_M=1017$ MeV,
$\beta_\rho=340$ MeV.}


\begin{tabular}{cccc}
\hline\hline
Integral & $\pi$-meson  & K-meson & $\eta$-meson \\
\hline
$I_1$ & 0.108 & 0.045 & 0.040 \\
$I_2$ & 0.113 & 0.058 & 0.053 \\
$I_3$ & 0.059 & 0.027 & 0.024 \\
$I_4$ & 0.105 & 0.024 & 0.020 \\
$I_5$ & 0.108 & 0.045 & 0.040 \\
$I_6$ & 0.126 & 0.087 & 0.082 \\
$I_7$ & 0.100 & 0.073 & 0.069 \\
$I_8$ & 0.057 & 0.027 & 0.024 \\
$I_9$ & 0.005 & 0.002 & 0.002 \\
\hline \hline
$\mathcal{I}_1$ & 0.008 & 0.025 & 0.024 \\
$\mathcal{I}_2$ & 0.038 & 0.016 & 0.014 \\
\hline \hline
\end{tabular}
\label{table:integrals}
\end{table}

Since our loop calculations involve divergent integrals,
cutoff-dependence of the results is inevitable. To explore this
point further, we also estimate the integrals using a monopole
form for the meson form factors suggested by the vector dominance
model \cite{VDM_FFs}. The cut-off parameters $\lambda$'s for the
monopole form factors are chosen such that the mean square radii
determined by the two forms (monopole and dipole) are the same. We
show in Table\ \ref{table:integrals_mono} the numerical values of
$I_i$ $(i=1, ..., 9)$, $\mathcal{I}_1$, and $\mathcal{I}_2$ for
different meson loops for case of using monopole form factors. All
the integrals are measured in MeV. Calculations use
$\lambda_A=601$ MeV, $\lambda_Q=596$ MeV, $\lambda_M=719$ MeV,
$\alpha_{\textrm{em}}=1/137$, $f \approx 93.0$ MeV, $\beta=0.75$,
and $\beta_\rho=340$ MeV.

\begin{table}
\caption{Numerical values of $I_i$ $(i=1, ..., 9)$,
$\mathcal{I}_1$, and $\mathcal{I}_2$ for different meson loops
when using monopole form factors. All the integrals are measured
in MeV. Calculations use $\alpha_{\textrm{em}}=1/137$, $f \approx
93.0$ MeV, $\beta=0.75$, $\beta_\rho=340$ MeV, $\lambda_A=601$
MeV, $\lambda_Q=596$ MeV, and $\lambda_M=719$ MeV.}


\begin{tabular}{cccc}
\hline\hline
Integral & $\pi$-meson  & K-meson & $\eta$-meson \\
\hline
$I_1$ & 0.176 & 0.097 & 0.090 \\
$I_2$ & 0.185 & 0.110 & 0.102 \\
$I_3$ & 0.113 & 0.069 & 0.065 \\
$I_4$ & 0.132 & 0.041 & 0.035 \\
$I_5$ & 0.176 & 0.097 & 0.090 \\
$I_6$ & 0.194 & 0.149 & 0.144 \\
$I_7$ & 0.168 & 0.135 & 0.131 \\
$I_8$ & 0.111 & 0.069 & 0.064 \\
$I_9$ & 0.008 & 0.004 & 0.004 \\
\hline \hline
$\mathcal{I}_1$ & 0.052 & 0.060 & 0.058 \\
$\mathcal{I}_2$ & 0.045 & 0.021 & 0.020 \\
\hline \hline
\end{tabular}
\label{table:integrals_mono}
\end{table}

\section{The coefficients $a$, $b$, $c$, and $d$}


To evaluate the coefficients $a$, $b$, $c$, and $d$, we first need
to estimate the coefficients $a_{\textrm{CM}}$, $b_{\textrm{CM}}$,
$c_{\textrm{CM}}$, and $d_{\textrm{CM}}$ given by Eq.
(\ref{cm-contrib}). Using $m_u \approx 340$ MeV, $(m_d-m_u)\approx
2.5$ MeV, $(m_d-m_u)/m_u =0.0074 \pm 0.0012 $, and a result from
the simple quark model \footnote{See, for example, the Halzen \&
Martin book \cite{Halzen-Martin} p.66.}
\begin{equation}
6 \frac{4\pi\alpha_s}{9}A = m_\Delta - m_N \approx 300 \,
{\textrm{MeV}} \, ,
\end{equation}
we get $C = 0.37 \pm 0.06$ MeV. Thus, for $m_s \approx 500$ MeV,
the color-magnetic contributions to the four parameters $a$, $b$,
$c$, and $d$ (in MeV) are
\begin{equation}\label{cm_cont1}
a_{\textrm{CM}}=-0.74 \pm 0.12, \, \, b_{\textrm{CM}}=0.75 \pm
0.12, \, \, c_{\textrm{CM}}=-0.75 \pm 0.12, \, \,
d_{\textrm{CM}}=0.70 \pm 0.44 \, .
\end{equation}
Next, using $\Delta^M_{\textrm{em}} = 1260$ ${\textrm{MeV}}^2$,
$\Delta^M_q = 5196$ ${\textrm{MeV}}^2$, and the values of the
integrals given in Table\ \ref{table:integrals} (Table\
\ref{table:integrals_mono} for the case of using monopole form
factors), it is straightforward to calculate numerically the
coefficients $b$, $c$, and $d$. To evaluate the coefficient $a$,
we have used $\Delta_{du}$ as a parameter and fitted $a$ exactly
to its experimental value. A best fit is obtained at
$\Delta_{du}=2.54 \pm 0.00$ MeV ($\Delta_{du}=2.43 \pm 0.00$ MeV
when using monopole form factors), a value that is lower than the
value $\Delta_{du} \approx 4.04$ MeV estimated ignoring loop
corrections to $\tilde{\alpha}_m$ \footnote{It is, however, close
to the Franklin and Lichtenberg' s value $\Delta_{du}=2.84 \pm
0.09 $ MeV found using the baryon mass differences \cite{FL1}.}.
The calculated values of the coefficients $a$, $b$, $c$, and $d$
(in MeV) are
\begin{equation}
\label{coeffs_case1} a=1.82 \pm 0.12, \, \, b=3.00 \pm 0.12, \, \,
c=-1.46 \pm 0.12, \, \, d=2.28 \pm 0.44 \, ,
\end{equation}
for the case of using dipole form factors, and
\begin{equation}
\label{coeffs_case2} a=1.82 \pm 0.12, \, \, b=3.34 \pm 0.12, \, \,
c=-1.50 \pm 0.12, \, \, d=2.48 \pm 0.44 \, ,
\end{equation}
for the case of using monopole form factors.

The calculated values of $a$, $b$, $c$, and $d$ for
$\mathcal{H}_{\textrm{IB}}$ and the contributions to those values
from $\mathcal{H}_1$, $\mathcal{H}_2$, $\mathcal{H}_3$, and
$\mathcal{H}_{\textrm{CM}}$ for these two cases are shown in
Table\ \ref{table:abcd}. The best fit values of the parameters
from $\mathcal{H}_{\textrm{IB}}^{\textrm{best}}$ are also given
there.


\begin{table}
\caption{Coefficients $a$, $b$, $c$, and $d$ from $\mathcal{H}_1$,
$\mathcal{H}_2$, $\mathcal{H}_3$, $\mathcal{H}_{\textrm{CM}}$,
$\mathcal{H}_{\textrm{IB}}$, and
$\mathcal{H}_{\textrm{IB}}^{\textrm{best}}$. All values of the
coefficients are in MeV. Calculations use $\Delta_{em}^{M}=1260$
${\textrm{MeV}}^2$, $\Delta_{q}^{M}=5196$ ${\textrm{MeV}}^2$, the
calculated values of the color-magnetic parameters given in Eq.
(\ref{cm_cont1}), $\Delta_{du}=2.54 \pm 0.00$ MeV  obtained from a
fit, and the values of the integrals shown in Table
\ref{table:integrals} ($\Delta_{du}=2.43 \pm 0.00$ MeV and Table
\ref{table:integrals_mono} for the case of using monopole form
factors). The uncertainties shown for the theoretical values of
the coefficients include only those from the uncertainties in the
color-magnetic terms.}

\begin{tabular}{c|rrrr|rrrr}
\hline\hline Hamiltonian & \multicolumn{4}{c|}{Dipole case} &
\multicolumn{4}{c}{Monopole case} \\
\cline{2-9}
& $a$  & $b$ & $c$ & $d$ & $a$  & $b$ & $c$ & $d$ \\
\hline
$\mathcal{H}_1$ & $-$0.05 & 2.17 & $-$0.08 & 0.74 & 0.01 & 2.29 & $-$0.18 & 0.67 \\
$\mathcal{H}_2$ & 0.07 & $-$0.12 & $-$0.43 & 0.04 & 0.12 & $-$0.03 & $-$0.23 & $-$0.23 \\
$\mathcal{H}_3$ & 2.54 & 0.20 & $-$0.20 & 0.80 & 2.43 & 0.33 & $-$0.33 & 1.34 \\
$\mathcal{H}_{\textrm{CM}}$ & $-$0.74 & 0.75 & $-$0.75 & 0.70 & $-$0.74 & 0.75 & $-$0.75 & 0.70 \\
\hline $\mathcal{H}_{\textrm{IB}}$ & 1.82 $\pm$ 0.12 & 3.00 $\pm$
0.12 & $-$1.46 $\pm$ 0.12 & 2.28 $\pm$ 0.44 & 1.82 $\pm$ 0.12 &
3.34 $\pm$ 0.12 & $-$1.50 $\pm$ 0.12 & 2.48 $\pm$ 0.44 \\
$\mathcal{H}_{\textrm{IB}}^{\textrm{best}}$ & 1.82 $\pm$ 0.04 &
3.35 $\pm$ 0.24 & $-$1.78 $\pm$ 0.23 & 1.00 $\pm$ 1.40
& 1.82 $\pm$ 0.04 & 3.35 $\pm$ 0.24 & $-$1.78 $\pm$ 0.23 & 1.00 $\pm$ 1.40 \\
\hline \hline
\end{tabular}
\label{table:abcd}
\end{table}

The results given in Table \ \ref{table:abcd} show that the
calculated IB corrections are of the right general size and have
the correct sign pattern. These corrections seem to account fairly
well for the pattern of coefficients $a$, $b$, $c$ in the IB
mass-difference Hamiltonian.  The coefficient $d$ has the correct
sign, but its calculated magnitude appears to be somewhat large.
However, like the other calculated values of $a$, $b$, and $c$,
the calculated value of $d$ agrees with its experimental value
within the margin of error. Using the values of $a$, $b$, $c$, and
$d$ given in Eq. (\ref{coeffs_case1}) for the dipole case, we can
easily evaluate the baryon mass splittings shown in Eq.
(\ref{splittings}) and find that the average deviation of the
calculated values from experiment is $0.23$ MeV with a weighted
$\chi^2/n_f$ of $1.63$ for $n_f=3$ degrees of freedom ($p\approx
0.2$).


Note that employing a monopole form for the meson form factors
improve the results for the coefficients $b$ and $c$, but not for
$d$. Nevertheless, the calculated value of $d$ still agrees with
its experimental value within the margin of error. Using the
values $a$, $b$, $c$, and $d$ given in Eq. (\ref{coeffs_case2})
for the monopole case, we find the average deviation of the
calculated values from experiment to be $0.29$ MeV with a weighted
$\chi^2/n_f=0.86$ for $n_f=3$ degrees of freedom ($p\approx
0.36$). The weighted fit is clearly better even though the average
deviation is larger, the result of the uncertainty in the
experimental value of $d$.

Since most of the integrals clearly have potentially large
short-distance contributions (the divergences in the absence of
form factors are strong), the presence of the short-distance
effects is likely the source of any deviations from the experiment
results. We do not regard the distinction between the dipole and
monopole fits as definitive given the different short-distance
behavior of the corresponding integrals.

To illustrate the uncertainty within the model, we note that if
one supposes that the quark wave functions are exponential at
short distances rather than Gaussian, $\exp{(-r/a)}$ rather than
$\exp{(-\beta_\rho^2 r^2/2)}$, but keep the dependence on
$\lambda$ and the mean square separation $<r^2>$ the same, then
$a=1/(\sqrt{2}\beta_\rho)$ and the Coulomb integrals increase by
$\sqrt{\pi/2}$. That is, the Coulomb integrals are underestimated
using Gaussian wave functions which neglect short distance
correlations. The more complicated wave functions used by Carlson
{\it et al.} \cite{Carlson} include correlations and indeed give a
some what larger energy as noted in \cite{DH05}. The same remarks
hold for the magnetic integral. The short distance effects are
even stronger there: the magnetic integrals calculated using the
exponential wave functions increase by a factor $2\sqrt{2\pi}$.
Therefore, $I_{QQ}$ and $I_{\mu\mu}$ can be treated as parameters
by multiplying their expressions defined in Eq. (\ref{IqqImm})
with the scale factors. We find, however, that doing so does not
substantially improve the results.

It is encouraging that the model, with all parameters except
$\Delta_{du}$ fixed gives as an accurate account of the purely
electromagnetic parameters $b$, $c$, $d$.


\section{\label{sec:conclusions} Conclusion}

Our results here consist of a numerical analysis of the IB
contributions to the baryon masses which we analyzed recently
using a loop expansion in the heavy baryon approximation to chiral
effective field theory and methods developed in our earlier
analyses of the baryon masses and magnetic moments. The leading IB
corrections to the baryon masses come from the Coulomb
interaction, magnetic moments interactions, the effects of the
$d,\,u$ quark mass difference, and effective point interactions
attributable to color-magnetic effects.

We have made reasonable estimates of the various integrals by
introducing the known form factors and wave functions from
successful semirelativistic models for the baryons and using the
results to evaluate the parameters. We find that the resulting
contributions to the IB corrections are of the right general size,
have the correct sign pattern, and agree with the experimental
values within the margin of error. We also find that employing
monopole instead of dipole form factors slightly improve the
results, but scaling the Coulomb or magnetic moment contributions,
does not. To the extent to which effects of adding a second meson
loop are small, a view supported by the smallness of the
three-body violations of the sum rules, it appears likely that any
deviations from the experiment can be attributed to the presence
of short-distance effects which can be parametrized but not
calculated in the effective field theory.

\appendix *
\section{One- and two-body operators} \label{1-2-operators}

We present here sets of one- and two- body operators defined
earlier in \cite{Morpurgo_EM1, DH05}. In the case of
$\textrm{O}(e^2)$ contributions to the baryon masses, the
$\Gamma$'s must be bilinear in the quark charge matrix $Q =
\textrm{diag}\,(2/3,-1/3,-1/3)$ and can depend otherwise on the
quark spin matrices $\bsig$ and and flavors. Ignoring isospin
breaking through the small $u$, $d$ mass difference and using the
conventions that
\begin{equation}
\label{sums} \sum[i]\equiv\sum_{i},\quad
\sum[ij]\equiv\frac{1}{2}\sum_{i\not=j}, \quad
\sum[ijk]\equiv\frac{1}{6} \sum_{i\not=j\not=k},
\end{equation}
where $ i,j,k\in u,d,s$ label the three quarks in a baryon, we
can group the $\Gamma$'s into sets of one- and  two-quark
operators as follows:
\begin{eqnarray}
&& \textrm{One-body operators:} \nonumber \\
\label{1,7}
&& \Gamma_1=\sum[Q_i^2],\quad \Gamma_7=\sum[Q_i^2M^s_i].
\end{eqnarray}
\begin{eqnarray}
&& \textrm{Two-body operators:} \nonumber \\
\label{2,4,5}
&& \Gamma_2=\sum[Q_i^2(\bsig_i\cdot\bsig_j)], \quad \Gamma_4=\sum[Q_iQ_j],\quad \Gamma_5 = \sum[Q_iQ_j(\bsig_i\cdot\bsig_j)], \\
\label{8,10,11}
&& \Gamma_8=\sum[Q_i^2M^s_i(\bsig_i\cdot\bsig_j)] ,\quad \Gamma_{10}=\sum[Q_i^2M^s_j],\quad \Gamma_{11}= \sum[Q_i^2M^s_j(\bsig_i\cdot\bsig_j)] ,\\
\label{13,14}
&& \Gamma_{13}=\sum[Q_iQ_jM^s_i], \quad \Gamma_{14}=\sum[Q_iQ_jM^s_i(\bsig_i\cdot\bsig_j)],   \\
\label{19,20}
&& \Gamma_{19}=\sum[Q_i^2M^s_iM^s_j], \quad \Gamma_{20}=\sum[Q_i^2M^s_iM^s_j(\bsig_i\cdot\bsig_j)], \\
\label{25,26}
&& \Gamma_{25}=\sum[Q_iQ_jM^s_iM^s_j], \quad
 \Gamma_{26}=\sum[Q_iQ_jM^s_iM^s_j(\bsig_i\cdot\bsig_j)] \, .
\end{eqnarray}

\begin{acknowledgments}

The author would like to thank Professor Loyal Durand for his
useful comments and invaluable support, and Professor Jerrold
Franklin for pointing out the omission of the color magnetic terms
in the original version of this paper, and for other useful
comments.

\end{acknowledgments}

\bibliography{EM_Bib}

\end{document}